\documentclass{ifacconf}

\usepackage{graphicx}      % include this line if your document contains figures
\usepackage{natbib}        % required for bibliography
\usepackage[dvips]{epsfig}    % or this line, depending on which
\usepackage{xcolor}

\usepackage{amssymb}
\usepackage{amsmath}          % Include this line if your
\usepackage{enumerate}

\begin{document}
\begin{frontmatter}

\title{Power based adaptive compensator of output oscillations}

\thanks[]{\textcolor[rgb]{0.00,0.00,1.00}{Author's accepted manuscript (IFAC MICNON 2024)}}

\author{Michael Ruderman}
\address{University of Agder \\ Department of Engineering Sciences \\ 4879-Grimstad, Norway  \\
Email: \tt\small michael.ruderman@uia.no}

%%%%%%%%%%%%%%%%%%%%%%%%%%%%%%%%%%%%%%%%%%%%%%%%%%%%%%%%%%%%%%%%%%%%%%%%%%%%%%%%
\begin{abstract}
Power-based output feedback compensator for oscillatory systems is
proposed. The average input-output power of an oscillatory signal
serves as an equivalent control effort, while the unknown
amplitude and frequency of oscillations are detected at each
half-period. This makes the compensator adaptive and discrete,
while the measured oscillatory output is the single available
signal in use. The resulting discrete control scheme enables a
drastic reduction of communication efforts in the control loop.
The compensator is designed for 2nd order systems, while an
extension to higher-order dynamics, like e.g. in case of
two-inertia systems, is also provided. Illustrative experimental
case study of the 5th order oscillatory system is provided.
\end{abstract}

\begin{keyword}
Power based control \sep oscillatory systems \sep harmonics
compensation \sep peak detection
\end{keyword}

\end{frontmatter}

%===============================================================================

%%%%%%%%%%%%%%%%%%%%%%%%%%%%%%%%%%%%%%%%%%%%%%%%%%%%%%%%%%%%%%%%%%%%%%%%%%%%%%%%
\section{Introduction}
\label{sec:1}

Control of the oscillating outputs in various types of the systems
is relevant in different applications. One can find those in e.g.
active and flexible structures (\cite{preumont2018vibration}),
robotics with elastic elements (\cite{DeLuca2016robots}), power
electronics, for instance inverters (\cite{wu2017damping}),
suspension systems (\cite{tseng2015state}), to mention here the
few. For a control-based rejection (or at lest attenuation) of the
oscillatory disturbances, an adaptation or online estimation of
the dynamic state/-s (usually more than one) can be required, see
e.g. \cite{DeWit2000adaptive}, \cite{aranovskiy2013adaptive},
\cite{ruderman2024adaptive}, and \cite{landau2020use} with
references therein. Often, also a robust online estimation, see
e.g. \cite{hsu1999globally}, \cite{bobtsov2012switched},
\cite{ruderman2022}, of the corresponding oscillation frequency,
which can be unknown or uncertain, is also required. When an
estimation of the required parameters, correspondingly states,
become a part of the compensation scheme itself, an often
nontrivial proof of stability can arise as strictly necessary.
Consequently, the overall control of oscillations can be sensitive
to the noise of output measurements, uncertain system parameters,
and phase lag (correspondingly delay) associated with a possible
use of the necessary additional filters.

Another way of looking on systems, that is also motivating the
present work, is based on an energy-, correspondingly
power-balancing, see e.g. \cite{garcia2010power}. For an
input-output system representation, a power-flow and power-shaping
represent an easily interpretable and mathematically elegant way
to analysis and control synthesis. The basics of energy shaping
principles in control can be looked in e.g. lecture notes of
\cite{ortega2001putting}. It is worth emphasizing, at that point,
that the oscillatory quantities are particularly straightforward
for a power-based representation, correspondingly calculations.

While most of the power- and energy-based control methods imply
the continuous control laws and, therefore, assume a one-to-one
correspondence (respectively mapping) between a vector of the used
system states and the control variables, another argument equally
supports the approach proposed in this paper. The set goal of
compensating for a particular oscillation quantity makes it
possible to reduce significantly the commutation rate of the
controller and enables a feedback control logic that commutates
for only few fractions of the oscillation period. This yields the
proposed compensation scheme to some type of an event-triggered
feedback control, see e.g. \cite{heemels2012introduction} and
references therein. For the related basics on hybrid (i.e.
continuous and switching) control systems we also refer to
\cite{liberzon2003} and \cite{lunze2009handbook}. Examples of the
hybrid event-switching controls are well known, for instance a
bang-bang funnel controller, \cite{liberzon2013bang}, and
event-triggered controller for saturated linear systems,
\cite{seuret2016lq}.

Against the above background, a novel discrete-valued power-based
adaptive control of the output oscillations is proposed. The main
advantage of the proposed control scheme, over other output
feedback based oscillation compensation approaches, is a largely
reduced communication effort in case of a digitally connected
system framework, i.e. remote location of the sensing and control
(or actuating) elements. In fact, any continuous feedback control
requires a communication effort of $\: 2 \pi \omega^{-1} f_s
\times const \:$ per oscillations period, where $\omega$ is the
angular frequency of oscillations and, most importantly, $f_s$ is
the digital sampling frequency. On the contrary, the proposed
feedback compensation has a communication effort of $\: 4 \pi
\omega^{-1} \times const$, since assigning the updated control
values only twice per oscillations period. Obviously, for some
large $f_s$ values (like kHz range or even larger) such reduction
of the communication effort can be beneficial in multiple regards.

The rest of the paper is organized as follows. The main results
are given in section \ref{sec:2}. First, the power-based control
is developed for second-order systems using the input-output power
balance of oscillatory signals. Then, the proposed control is
extended for higher-order systems by considering the propagation
of compensating signal through the forward dynamics. An extrema
detection algorithm, which constitutes an inherent part of the
proposed control scheme, is also given in detail. An experimental
case study of compensating oscillations in the fifth-order system,
with two-masses connected by a low-damped spring, is provided in
section \ref{sec:3}. The case of an additional excitation of the
oscillating behavior by the external (mechanical) disturbances is
also shown, in favor of the robustness of the proposed control
scheme. Short summary is given by the end in section \ref{sec:4}.

%%%%%%%%%%%%%%%%%%%%%%%%%%%%%%%%%%%%%%%%%%%%%%%%%%%%%%%%%%%%%%%%%%%%%%%%%%%%%%%%
\section{Main results}
\label{sec:2}

\subsection{Power-based compensator} \label{sec:2:sub:1}

\begin{figure}[!h]
\centering
\includegraphics[width=0.75\columnwidth]{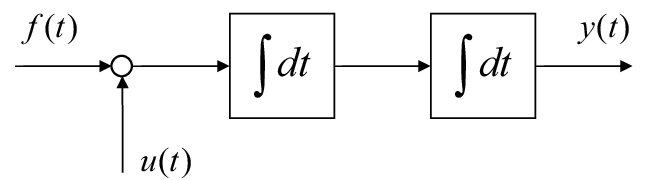}
\caption{Signals flow in 2nd order system with output
\eqref{eq:2:1:1}.} \label{fig:2ndorder}
\end{figure}

Consider next the class of second-order systems with the
measurable oscillatory output
\begin{equation}
y(t) = A \sin (\omega t + \varphi) + \Psi. \label{eq:2:1:1}
\end{equation}
Assume the oscillations amplitude and frequency $A,\omega > 0$ are
uncertain, and only the frequency upper bound $\Omega_{\max} >
\omega$ is known. The phase shift $\varphi$ is insignificant, as
shown later, especially since the formulated power-based control
operates on every period, and $y(t)$ is available. A
non-oscillatory term $\Psi$ is assumed to be known and $|d\Psi/dt|
\ll A \omega$, i.e. the oscillation dynamics in $\dot{y}(t)$
dominates over the dynamics of $\Psi$. Moreover, the oscillations
amplitude itself can be a slowly varying process; we notice that
'slow' here is also in comparison with the oscillations dynamics
i.e. $|dA/dt| \ll A \omega$. Therefore, in the following
developments, $A$ will be considered as a 'frozen' (to say
quasi-constant) process parameter, at least over one period, while
the resulted control behavior yields effective for a slowly
changing $A(t)$ as well. The system with output \eqref{eq:2:1:1},
which is equal to
$$
y(t) = \iint \bigl( u(t) + f(t) \bigr) dt,
$$
has a structure depicted in Fig. \ref{fig:2ndorder}, while the
control input $u(t)$ is available for compensating the
oscillations in $y(t)$. Recall that the driving input $f(t)$ is
unknown.

Differentiating twice the signal \eqref{eq:2:1:1} we obtain
\begin{equation}
\ddot{y}(t) = - A \omega^2  \sin (\omega t + \varphi) +
\ddot{\Psi}. \label{eq:2:1:2}
\end{equation}
Denoting the oscillatory part of \eqref{eq:2:1:1} by $\tilde{y}$,
we obtain the corresponding part of the (instantaneous) power flow
as
\begin{equation}
\tilde{P}(t) = \ddot{\tilde{y}}(t) \, \tilde{y}(t) = - A^2
\omega^2 \sin (\omega t + \varphi)^2. \label{eq:2:1:3}
\end{equation}
Note that $\ddot{\Psi} \approx 0$ can be assumed for oscillatory
part without loss of generality. It is obvious that for one
period, equally as for an average with $t \rightarrow \infty$, the
input-output power flow yields
\begin{equation}
P =  - \frac{1}{2} A^2 \omega^2. \label{eq:2:1:4}
\end{equation}
In order to compensate for \eqref{eq:2:1:4}, the corresponding
control power $P^* = -P$ is required, and an equivalent control
effort (per one period) can be obtained as
$$
U \propto \frac{|P|}{A} = \frac{1}{2} \, \omega^2 A.
$$
When using an oscillations suppressing control $u(t)$, the
corresponding amplitude $A$ and so the input-output power
\eqref{eq:2:1:4} reduces during the operation. Thus, the control
power $P^*(\tilde{A},\tilde{\omega})$ needs an instantaneous
amplitude $\tilde{A}$ to be estimated, but also an instantaneous
frequency estimate $\tilde{\omega}$; the latter due to $\omega$ is
assumed to be uncertain to some extent. Since both are detectable
between two successive extrema in $y(t)$, an estimate
(correspondingly update) of the $(\tilde{A},\tilde{\omega})$ pair
appears twice per period.

Denoting by $t^*$ the time instant of the last extrema, i.e.
either minimum of maximum of the oscillating $y(t)$, the proposed
power-based control has the form
\begin{equation}
u(t) = K \, \tilde{\omega}^2 \tilde{A}(t^*) \label{eq:2:1:5}
\end{equation}
with
$$
\tilde{A}(t^*) \equiv \tilde{A} \, \mathrm{sign} \bigl( y(t^*) -
\Psi(t^*) \bigr).
$$
We next determine an optimal (over one period) gain $K > 0$, while
emphasizing that the control \eqref{eq:2:1:5} keeps a constant
value between two successive extrema.

While the input-output power of the oscillatory output, which has
to be compensated, is given by \eqref{eq:2:1:3}, the input-output
power of a constant input $u = U$ (if zeroing the
oscillations-driving input $f=0$) is given by
\begin{equation}
P^* = U  \iint U dt = \frac{1}{2} \, U^2 t^2. \label{eq:2:1:6}
\end{equation}
Integrating \eqref{eq:2:1:3} and \eqref{eq:2:1:6} over one period,
we obtain the corresponding energies that can be balanced as
\begin{equation}
\int \limits_{0}^{2\pi / \omega} \bigl( - A^2 \omega^2 \sin
(\omega t + \varphi)^2 \bigr) dt + \frac{1}{2} \int
\limits_{0}^{2\pi / \omega} U^2 t^2 dt = 0. \label{eq:2:1:7}
\end{equation}
Solving \eqref{eq:2:1:7} with respect to $U$ we obtain
$$
U = \frac{\sqrt{3}}{2\pi} \, \omega^2 A,
$$
cf. with the control \eqref{eq:2:1:5}. Following to that, an
optimal gain for compensating the oscillations in $y(t)$ with
\eqref{eq:2:1:5} is
\begin{equation}
K = \frac{\sqrt{3}}{2\pi}. \label{eq:2:1:8}
\end{equation}
Note that the determined gain \eqref{eq:2:1:8} is rather
conservative. Indeed, the energy balance \eqref{eq:2:1:7} is
considered over one full period, while the extrema detection and,
correspondingly, update of the oscillation parameters in
\eqref{eq:2:1:5} take place every half-period. However, the
balancing of input-output power (respectively of energy) turns out
more reasonable over a full period $[t, \, 2\pi / \omega +t]$, due
to a possible time delay of the extrema detection, cf. later in
section \ref{sec:2:sub:3}. This allows avoiding an
overcompensation which, otherwise, can additionally excite the
oscillations in $y(t)$.

\subsubsection{Numerical example:} \label{sec:2:sub:1:subsub}

\begin{figure}[!h]
\centering
\includegraphics[width=0.49\columnwidth]{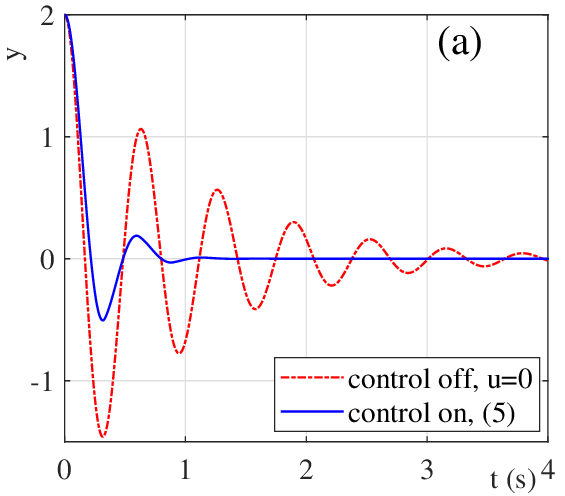}
\includegraphics[width=0.49\columnwidth]{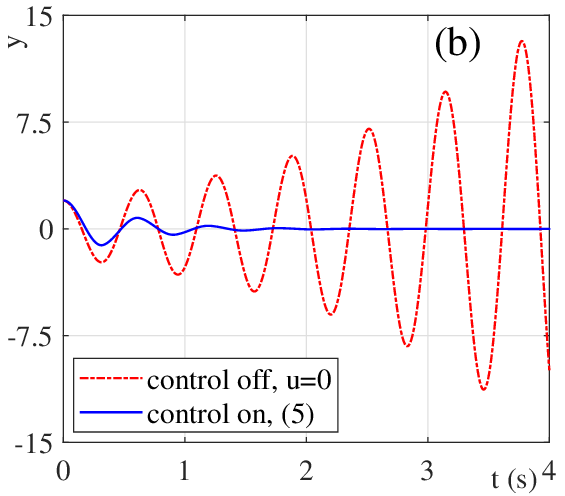}
\includegraphics[width=0.49\columnwidth]{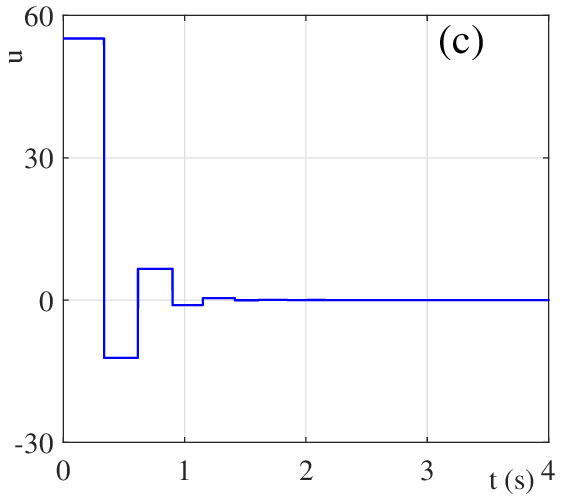}
\includegraphics[width=0.49\columnwidth]{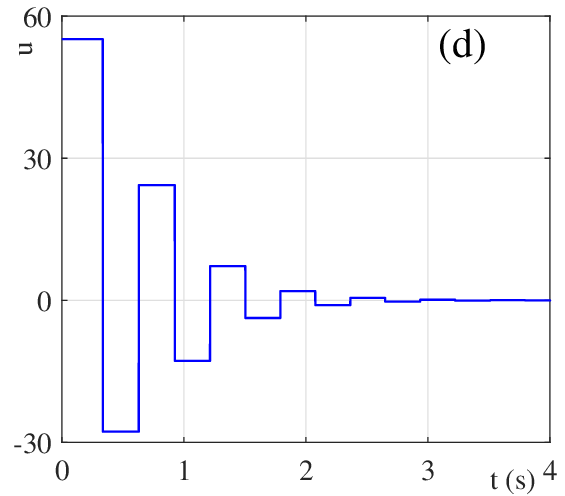}
\caption{Simulation results of the second-order oscillatory system
with and without use of compensator \eqref{eq:2:1:5}. The output
response and control value, in (a) and (c) for $a=2$, in (b) and
(d) for $a=-1$.} \label{fig:simres1}
\end{figure}

A second-order oscillatory system
\begin{equation}
\ddot{y}(t) + a \dot{y}(t) + b y(t) = u(t), \quad y(0) = c \neq 0
\label{eq:2:1:9}
\end{equation}
is used in the numerical simulation, once without feedback
compensator (i.e. $u=0$) and once using the power-based control
law \eqref{eq:2:1:5}. The assigned coefficient $b = 100$ results
in a natural frequency $\omega = 10$ rad/s. The initial value
constant is assigned to $c=2$. The damping coefficient is assigned
as $a = \{2, \, -1\}$, thus resulting in a homogenous solution of
\eqref{eq:2:1:9} which is oscillatory and exponentially converging
in the first case, and oscillatory and diverging in the second
case, respectively, see the red dashed lines in Fig.
\ref{fig:simres1} (a) and (b).
The compensated output response, i.e. when the control
\eqref{eq:2:1:5} is on, is shown by the blue solid line, while the
corresponding control values are depicted in Fig.
\ref{fig:simres1} (c) and (d), respectively.

\subsection{Extension to higher-order systems} \label{sec:2:sub:2}

\begin{figure}[!h]
\centering
\includegraphics[width=0.9\columnwidth]{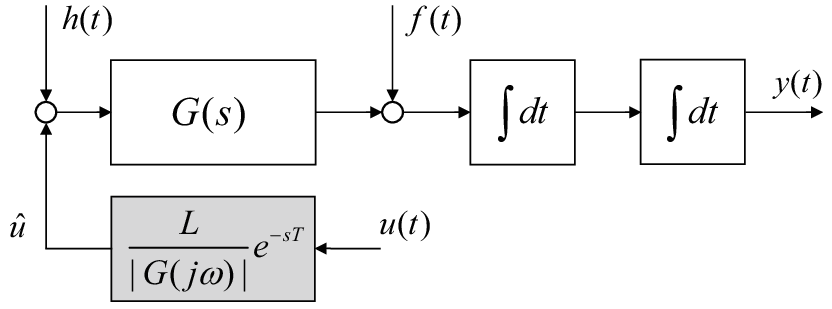}
\caption{Signals flow in higher-order system with output
\eqref{eq:2:1:1}.} \label{fig:HOsysten}
\end{figure}

For dynamic systems with order higher than two, the structure is
assumed as shown in Fig. \ref{fig:HOsysten}. The assumptions made
in section \ref{sec:2:sub:1} remain valid, while a rational stable
transfer function $G(s)$ is assumed to be known. The corresponding
output signal, expressed in Laplace domain, is then given by
$$
y(s) = \frac{1}{s^2} \biggl( G(s)\Bigl(h + \frac{L}{|G(j\omega)|}
\exp(-sT) u(s) \Bigr) + f \biggr).
$$
Furthermore, it is assumed that the unknown driving input signal
$h(t)$ provides as asymptotically stable response
$$
\int \limits_{0}^{t} g(t-\tau)h(t) d \tau,
$$
where $g(t)$ is the corresponding inverse Laplace transform of
$G(s)$, i.e. the impulse response function. Also assume that the
time constant(s) of $g(t)$ are significantly lower than $\pi
\omega^{-1}$, i.e. half-period of the output oscillations.

Since the compensator output $u(t)$, cf. \eqref{eq:2:1:5}, is no
longer matched with $f(t)$, cf. with Fig. \ref{fig:2ndorder}, its
propagation through $G(s)$ requires the following processing. The
available control channel $\hat{u}(t)$ of the system will be
modified by the amplitude response $|G(j \omega)|$ at the
characteristic frequency $\omega$. Moreover, an additional phase
lag $\arg \bigl[ G(j 2 \omega)\bigr]$ appears at twice of the
oscillation frequency. Here we recall that the control $u(t)$ is
of the discrete type and switches at half of the oscillation
period. Following to that, the compensator signal \eqref{eq:2:1:5}
undergoes the following transformation
\begin{equation}
\hat{u}(t) = L \, \Bigl|G(s)^{-1}\Bigr|_{s=j \omega} \, u(t-T).
\label{eq:2:2:1}
\end{equation}
The corresponding time delay factor
\begin{equation}
T = \Bigl(2\pi + \arg \bigl[ G(j 2 \omega)\bigr] \Bigr) \,
\omega^{-1} \label{eq:2:2:2}
\end{equation}
shifts the control value by the negative phase lag with respect to
a full period $2 \pi \omega^{-1}$. This synchronizes $u(t)$, which
is propagated through \eqref{eq:2:2:1}, again with the input of
double integrator, cf. Fig. \ref{fig:2ndorder}. Note, that the
rectangular pulse signal $u(t)$ is reshaped by $G(j\omega)$, thus
losing its energetic content, correspondingly for the impulse
$$
U \int \limits_0^{\pi / \omega} dt \:  > \int \limits_0^{\pi /
\omega} U  \int \limits_0^{t} g(t-\tau) d \tau dt.
$$
In order to regard for this inequality, a tunable impulse
weighting factor
\begin{equation}
1 < L < 3 \label{eq:2:2:3}
\end{equation}
is additionally used in \eqref{eq:2:2:1}.

\subsubsection{Numerical example:} \label{sec:2:sub:2:subsub}

\begin{figure}[!h]
\centering
\includegraphics[width=0.98\columnwidth]{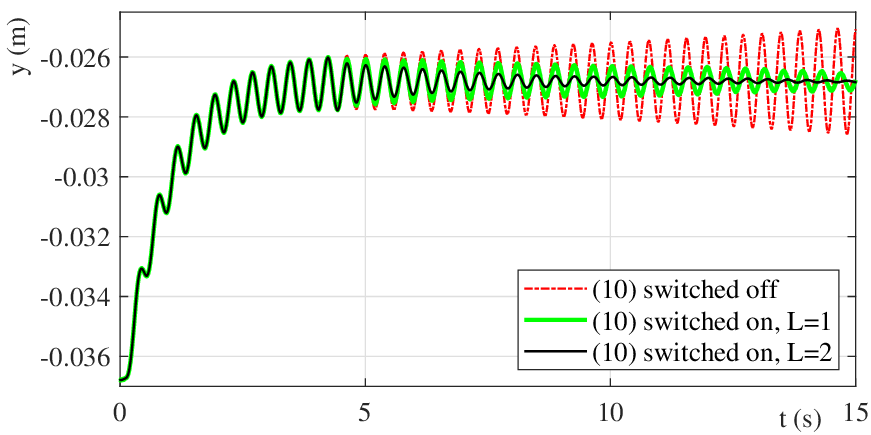}
\caption{Simulation results of the fifth-order oscillatory system
with and without use of compensator \eqref{eq:2:2:1}.}
\label{fig:simres2}
\end{figure}

A fifth-order oscillatory system, the same as shown later in the
experimental case study in section \ref{sec:3}, is numerically
simulated with the compensator \eqref{eq:2:2:1}. Note that for
keeping a known $\Psi$ value, cf. \eqref{eq:2:1:1}, a simple
proportional feedback control
$$
v(t) = 70\bigl(R_1 - y(t)\bigr) + R_2
$$
is additionally applied, which also leads to destabilization of
the closed-loop when not using \eqref{eq:2:2:1}. The first
constant $R_1$ constitutes a reference value. The second constant
$R_2$ is used to compensate for the total gravity term, cf. with
section \ref{sec:3:sub:1}. Further we note that for approaching
the real systems in a numerical simulation, the output $y(t)$ is
additionally subject to a band-limited white noise. The diverging
oscillatory behavior, when \eqref{eq:2:2:1} is switched off, is
shown in Fig. \ref{fig:simres2}.
When switching on the compensator \eqref{eq:2:2:1} at time $t=4$
sec, the otherwise unstable oscillations become stabilized, as
shown in Fig. \ref{fig:simres2} for two weighting factors $L=\{1,
\,2 \}$. Obviously $L$ affects the convergence rate of the
power-based oscillations compensation.

\subsection{Extrema detection} \label{sec:2:sub:3}

\begin{figure}[!h]
\centering
\includegraphics[width=0.98\columnwidth]{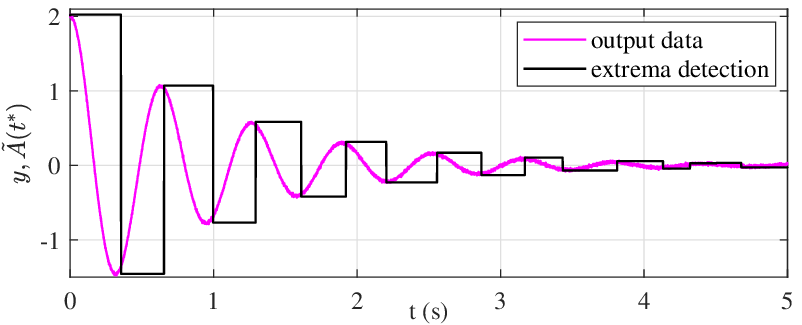}
\caption{Example of extrema detection from noisy output.}
\label{fig:extrema}
\end{figure}
An extrema detecting algorithm, required for $(\tilde{A},
\tilde{\omega})$ estimation at each $t^*_i$, cf. \eqref{eq:2:1:5},
is summarized below. The discrete time sampling index is $n$, the
size of the trapped delay buffer for smoothing filter is $N$, and
the index of recent extrema is $i$. An example of extrema
detection from the simulated $y(t)$, which is affected by the
band-limited white noise, is shown in Fig. \ref{fig:extrema} for
$f_s = 1$ kHz and $N=30$.
\begin{eqnarray*}
  1) & & \hbox{\texttt{Initialization}: } \tilde{A}(t^*_{i-1}) = y_0; \; \tilde{\omega}(t^*_{i-1}) = 0.5 \Omega_{\max}  \\[-0.5mm]
  2) & & \hbox{\texttt{For } } n=1 \hbox{\texttt{ to } } \infty \: \hbox{\texttt{ do } } \\[-0.5mm]
  3) & & \quad \bar{y}_n = \max \bigl(y_n, y_{n-1}, \ldots, y_{n-N} \bigr);  \\[-0.5mm]
  4) & & \quad S_n = \mathrm{sign}\bigl(\bar{y}_n - \bar{y}_{n-1}\bigr); \\[-0.5mm]
  5) & & \quad \hbox{\texttt{If } } S_n \neq 0 \, \wedge \, S_{n} \neq S_{i-1} \,  \wedge \, \pi/\bigl(n f_s - t^*_{i-1}\bigr) < \Omega_{\max}  \\[-0.5mm]
  6) & & \qquad \hbox{\texttt{Update}: }  \:S_i = S_n; \; t^*_i = n f_s;  \\[-0.5mm]
  7) & & \qquad \qquad \qquad \tilde{A}(t^*_{i}) = \bar{y}_n - \Psi; \; \tilde{\omega}(t^*_{i}) = \pi / \bigl(n f_s - t^*_{i-1}\bigr) \\[-0.5mm]
  8) & & \qquad \qquad \qquad i = 1+1; \\[-0.5mm]
  9) & & \quad \hbox{\texttt{End } } \\[-0.5mm]
  10) & & \hbox{\texttt{End } }
\end{eqnarray*}

%%%%%%%%%%%%%%%%%%%%%%%%%%%%%%%%%%%%%%%%%%%%%%%%%%%%%%%%%%%%%%%%%%%%%%%%%%%%%%%%
\section{Experimental case study}
\label{sec:3}

\subsection{Oscillatory system} \label{sec:3:sub:1}

\begin{figure}[!h]
\centering
\includegraphics[width=0.5\columnwidth]{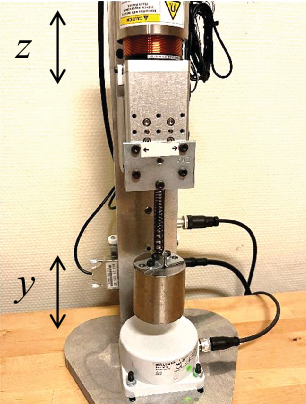}
\caption{Experimental setup of oscillatory system.}
\label{fig:exp}
\end{figure}

The following experimental case study is accomplished on the
two-mass oscillatory system under gravity, see
\cite{ruderman2021,ruderman2022,ruderman2024adaptive},
\cite{voss2022} for details. The setup shown in Fig. \ref{fig:exp}
consists of one free hanging load and one
linear actuator, based on the voice-coil-motor. Both moving bodies
with one vertical degree of freedom are connected via an elastic
spring with the relatively high stiffness and hardening effects
subject to unknown uncertainties. The actuator input voltage $v
\in [0,\, 10]$ V is the available control channel. The single
(contact-less) measured output value is the relative displacement
of the load $y(t)$. Note that the load is passive, while the
actuator displacement $z \in [0,\, 0.021]$ m remains an
unmeasurable state. Both the input and output values are real-time
available with the sampling rate set to $f_s = 5$ kHz.

The system structure is known, being given by
\begin{equation}
\rho(s) = \frac{3.2811}{0.0012 s + 1}\, v(s), \label{eq:5:1}
\end{equation}
\begin{eqnarray}
\label{eq:5:2}
  \dot{x}(t) &=& A \, x(t) + B \,\rho(t) + D, \\
\nonumber y(t) &=& C \, x(t),
\end{eqnarray}
with the state vector $x \in \mathbb{R}^4$ and
\begin{eqnarray}
\nonumber  A &=& \left(%
\begin{array}{cccc}
-333.35 &  -333.33  &   0.015     & 333.33  \\
1         & 0     &     0         & 0         \\
0.012     & 266.66 &    -0.012    & -266.66 \\
0         & 0          &  1       & 0
\end{array}%
\right), \\
\nonumber B &=& (1.667, 0, 0, 0)^\top, \quad  C \; = \; (0, 0, 0, 1), \; \hbox { and} \\
\nonumber D &=& (-9.806, 0, -9.806, 0)^\top.
\end{eqnarray}
At the same time, the identified parameters in \eqref{eq:5:1},
\eqref{eq:5:2} are not explicitly used for the control. Also worth
noting is that the disturbance $D$ is composed by the constant
gravity acting on both the moving actuator and load.

\begin{figure}[!h]
\centering
\includegraphics[width=0.98\columnwidth]{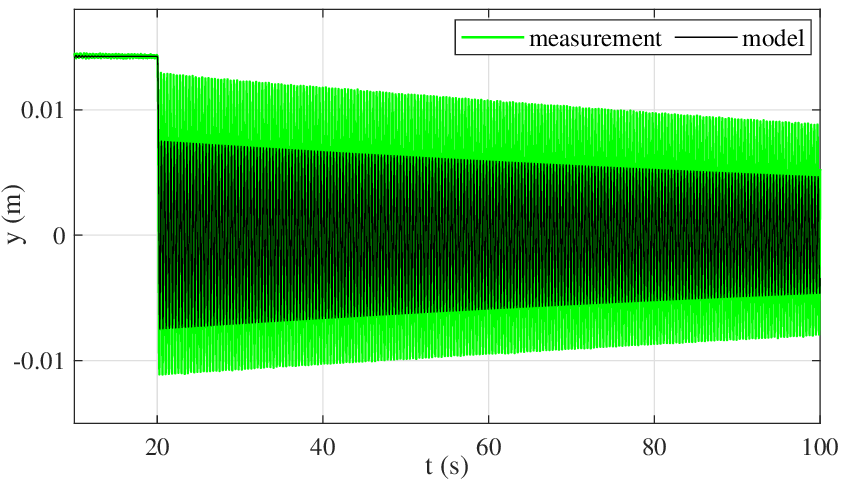}
\caption{Measured and computed oscillatory response.}
\label{fig:freefall}
\end{figure}

For exposing the low damped output oscillations, a free fall
scenario is performed, see Fig. \ref{fig:freefall}. Starting from
non-zeros initial conditions with $v(t) = \mathrm{const}$, which
compensates for the total gravity, the control signal is switched
off at $t=20$ sec. Due to $v(t>20)=0$ both moving masses fall
down, while $|\dot{y}|$ is larger than the actuator velocity due
to the bearing of the latter.
Once the actuator displacement experiences a hard impact at its
down mechanical limiter, the $y(t)$ becomes largely excited. It
begins to oscillate and is structurally damped by the spring only,
cf. Fig. \ref{fig:freefall}. Note that the oscillation amplitude
is sensitive to the initial conditions and exact knowledge of the
moving masses and stiffness. This leads to a visible amplitude
difference between the measured and computed output, while the
frequency and damping ratio match sufficiently well.

\subsection{Experimental control results} \label{sec:3:sub:2}

\begin{figure}[!h]
\centering
\includegraphics[width=0.98\columnwidth]{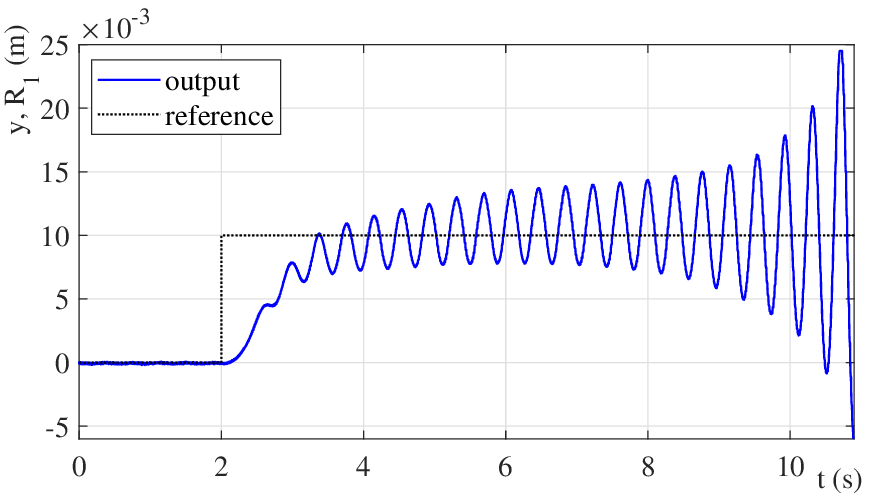}
\caption{Experimental results of PI-controlled load position
without use of the compensator \eqref{eq:2:2:1}.}
\label{fig:expres1}
\end{figure}

\begin{figure}[!h]
\centering
\includegraphics[width=0.98\columnwidth]{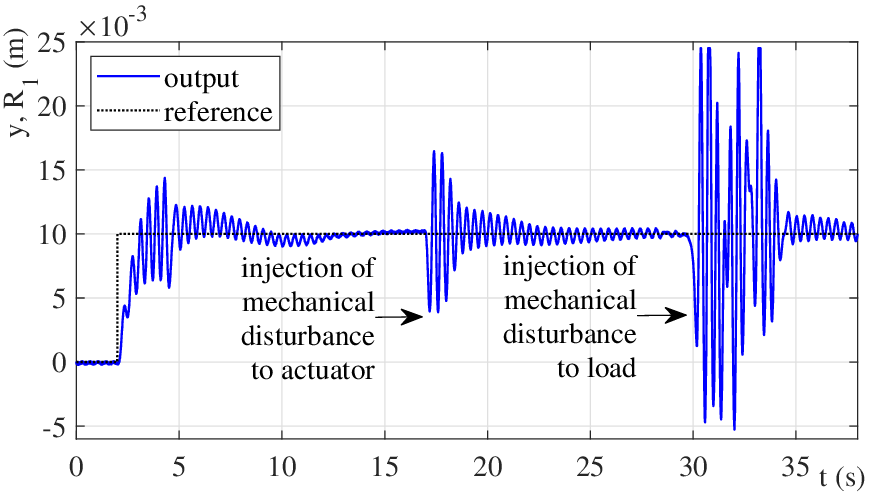}
\caption{Experimental results of PI-controlled load position with
use of the compensator \eqref{eq:2:2:1}. Mechanical disturbance is
injected once to actuator and once to load.} \label{fig:expres2}
\end{figure}

The experimental control results are obtained with the PI
(proportional-integral) output feedback control
$$
v(t) = 150 \bigl(R_1 - y(t)\bigr) + 170 \int \bigl(R_1 -
y(t)\bigr) dt + R_2 + \hat{u}(t),
$$
once without (i.e. $\hat{u}=0$) and once with the use of the
oscillations compensator \eqref{eq:2:2:1}. Note that a PI control
structure is necessary for keeping $\Psi \approx R_1$ despite all
additional uncertainties and disturbances (like for example
friction), which are not captured in \eqref{eq:5:1},
\eqref{eq:5:2}, cf. section \ref{sec:2:sub:2}. Here again, a
constant term $R_2$ is compensating for the total known gravity of
the actuator and load masses. An unstable diverging output
response for $\hat{u}=0$ is shown in Fig. \ref{fig:expres1}.
On the contrary, the compensator \eqref{eq:2:2:1}, switched on at
time $t=4$ sec, is suppressing the unstable oscillations, see Fig.
\ref{fig:expres2}. Moreover, an impulse-like external mechanical
disturbance was manually injected, once to the actuator (at time
around 17 sec), and once to the load (at time around 30 sec). Note
that the latter leads also to some vertical misalignments of the
hanging load and, in consequence, to an increase of the
measurement and process noise and, thus, less accurate extrema
detection. Still in both external disturbance cases, the output is
stabilized again after some transient oscillations.

%%%%%%%%%%%%%%%%%%%%%%%%%%%%%%%%%%%%%%%%%%%%%%%%%%%%%%%%%%%%%%%%%%%%%%%%%%%%%%%%
\section{Summary}
\label{sec:4}

A power-based discrete-valued adaptive compensator is proposed for
oscillatory outputs. Comparing to other approaches of compensating
the oscillations, like e.g. a most recent time-delay-based one
provided in \cite{ruderman23,ruderman2024adaptive}, the presented
control scheme commutates only twice per period of oscillations.
This offers a control efficiency in terms of the communication
efforts, significant in case of a digitally connected system
framework, i.e. when the output sensing and control
(correspondingly actuator) elements are remote. The proposed
control is derived and discussed for the second-order systems,
while an extension to higher-order systems is also given. The
proposed control requires a robust extrema detection, which allows
then also for operating with the noisy output signals. A
convincing experimental case study, accomplished on the
fifth-order oscillatory mechanical system, is also given.

%%%%%%%%%%%%%%%%%%%%%%%%%%%%%%%%%%%%%%%%%%%%%%%%%%%%%%%%%%%%%%%%%%%%%%%%%%%%%%%%
\section*{Acknowledgement}
\label{sec:5} This research was partially supported by the RCN
grant number 340782.

\bibliography{references}             % bib file to produce the bibliography
                                                     % with bibtex (preferred)

\end{document}